\documentclass[11pt,a4paper]{article}

\pdfoutput=1
\usepackage{jcappub}
\usepackage[latin1]{inputenc}
\usepackage{graphicx}
\usepackage{epsfig}
\usepackage{subfigure}
\usepackage{color}
\usepackage{comment}
\usepackage{slashed}

\def\beq{\begin{equation}}
\def\eeq{\end{equation}}
\def\baq{\begin{eqnarray}}
\def\eaq{\end{eqnarray}}

\newcommand{\be}{\begin{equation}} % only untightened
\newcommand{\ee}{\end{equation}}
\newcommand{\bea}{\begin{eqnarray}} % only untightened
\newcommand{\eea}{\end{eqnarray}}
\newcommand{\nn}{\nonumber}
\newcommand{\bmp}{\noindent\begin{minipage}{16cm}}
\newcommand{\emp}{\end{minipage}\vskip 7mm} % 7mm untightened
\def\lsim{\mathrel{\raise.3ex\hbox{$<$\kern-.75em\lower1ex\hbox{$\sim$}}}}
\def\gsim{\mathrel{\raise.3ex\hbox{$>$\kern-.75em\lower1ex\hbox{$\sim$}}}}

\newcommand{\intron}[1]{}%{#1}

\def\sfrac#1#2{{\textstyle\frac{#1}{#2}}}

\subheader{\small{Preprint: HIP-2016-19/TH}}

\title{A Strong Electroweak Phase Transition from the Inflaton Field}

\author[a,b]{Tommi Tenkanen,}
\author[a,b]{Kimmo Tuominen}
\author[b,c]{and Ville Vaskonen}

\affiliation[a]{Department of Physics, University of Helsinki \\
                      P.O.~Box 64, FI-00014, Helsinki, Finland}
\affiliation[b]{Helsinki Institute of Physics, \\
                      P.O.~Box 64, FI-00014, Helsinki, Finland}   
\affiliation[c]{Department of Physics, University of Jyvaskyla, \\
                      P.O.Box 35 (YFL), FI-40014 University of Jyvaskyla, Finland}
                            
\emailAdd{tommi.tenkanen@helsinki.fi}
\emailAdd{kimmo.i.tuominen@helsinki.fi}
\emailAdd{ville.vaskonen@jyu.fi}

\abstract{We study a singlet scalar extension of the Standard Model. The singlet scalar is coupled non-minimally to gravity and assumed to drive inflation, and also couple sufficiently strongly with the SM Higgs field in order to provide for a strong first order electroweak phase transition. Requiring the model to describe inflation successfully, be compatible with the LHC data, and yield a strong first order electroweak phase transition, we identify the regions of the parameter space where the model is viable. We also include a singlet fermion with scalar coupling to the singlet scalar to probe the sensitivity of the constraints on additional degrees of freedom and their couplings in the singlet sector. We also comment on the general feasibility of these fields to act as dark matter.
}


\begin{document}
\maketitle
%%%%%%%%%%%%%%%%%%%%%%%%%%%%%%%%%%%%%%%%%%%%%%%%%%%%%%%%%%%%%%%%%%%%%%%%%%%%%%%%%%%%%%%%%%%%%%%%%%%%

\section{Introduction}

Dynamics of cosmological inflation, mechanisms for creation of the observed dark matter abundance and the excess of matter over antimatter require non-minimal coupling to gravity and/or new degrees of freedom beyond the minimal Standard Model (SM) of elementary particle interactions. Many beyond the SM (BSM) theories address one or more of these issues, and a variety of different scenarios for each exist in literature, e.g. \cite{Bezrukov:2007ep, Burgess:2000yq,Ashoorioon:2009nf, Lerner:2009xg, Belyaev:2010kp, Cline:2012hg, Alanne:2014bra, Hambye:2008bq, DiChiara:2015bua}.

The common feature of most BSM constructions is the existence of a new scale between the Planck and the electroweak scales. The closer the new scale is to the electroweak scale, the better are the prospects for its testability at energies currently accessible at collider experiments. For example, of the several possibilities for baryogenesis, direct testability provides motivation for the idea of electroweak baryogenesis (EWBG), i.e. that the baryon asymmetry is generated during the electroweak phase transition. 

One of the necessary conditions for any successful baryogenesis scenario is out-of-equilibrium dynamics, which for the EWBG in particular is provided by a strongly first order electroweak phase transition. However, this requirement fails to be realized in the SM, where the phase transition is known to be a continuous cross-over \cite{Kajantie:1996mn}. Hence, one must look into BSM scenarios to make progress.

Even if the observed baryon asymmetry of our universe is produced by different mechanism than EWBG, the order of the electroweak phase transition is interesting also for the possibility to observe gravitational waves originating from a first order phase transition \cite{Hindmarsh:2015qta} in the early universe. The ground-based interferometer experiment LIGO has recently reported observations of gravitational waves from binary mergers and demonstrated the general feasibility of such observations \cite{Abbott:2016blz,TheLIGOScientific:2016pea}, and in the future space-based detectors like eLISA will have maximum sensitivity at the frequency range relevant for a first order phase transition at the 
electroweak scale \cite{Seoane:2013qna}.

In this paper we consider a singlet scalar field with a discrete $Z_2$ symmetry, and also the general case without this symmetry, and show that already in this minimal BSM scenario one can simultaneously address two important issues in the early universe physics: The singlet scalar, non-minimally coupled to gravity, is assumed to drive inflation, and also couple sufficiently strongly with the SM Higgs field in order to provide for a strong first order electroweak phase transition at a later stage in the history of the universe. 

Earlier studies on both inflation and strong EWPT have been conducted separately in this model in e.g. \cite{Lerner:2009xg} and \cite{Cline:2012hg, Alanne:2014bra}. Our new result is that we identify the regions of the parameter space where the model describes inflation successfully, is compatible with the LHC data, and yields a strong first order electroweak phase transition. We also include a singlet fermion coupled to the singlet scalar in order to probe the sensitivity of the constraints on additional degrees of freedom and their couplings in the singlet sector. These will eventually be necessary when extending the model to account for sufficient CP violation relevant for applications towards electroweak baryogenesis.

The paper is organized as follows: In section \ref{model} we define the model which we will study, and in sections \ref{cosmicinflation} and \ref{EWPT} we discuss, respectively, the details of cosmic inflation and properties of the electroweak phase transition in this model. In section \ref{conclusions} we conclude and discuss prospects for future work. 

\section{The Model}
\label{model}
The model we study is defined by the action
\be
\label{action}
S = \int d^4x \sqrt{-g}\bigg(\mathcal{L}_{\rm SM} + \mathcal{L}_{\rm gravity} - V(s,\phi) + \mathcal{L}_{\psi} \bigg),
\ee
where $\mathcal{L}_{\rm SM}$ is the Lagrangian density of the minimal SM, and the term $\mathcal{L}_{\rm gravity}$, whose exact form is specified in section \ref{cosmicinflation}, contains both the usual Einstein-Hilbert term and possible non-minimal couplings between the scalar fields and gravity. The potential
\be
\begin{aligned}
\label{scalarpotential}
V(s,\phi) =& \mu_{\rm \phi}^2 \phi^{\dagger}\phi + \lambda_{\rm h}(\phi^{\dagger}\phi)^2 + \mu_1^3 s + \frac{\mu_{\rm s}^2}{2}s^2 + \frac{\mu_3}{3}s^3 + \frac{\lambda_{\rm s}}{4}s^4 \\
&+ \mu_{\rm hs} (\phi^{\dagger}\phi)s + \frac{\lambda_{\rm hs}}{2}(\phi^{\dagger}\phi)s^2 ,
\end{aligned}
\ee
is the most general renormalizable scalar potential in a Higgs portal model where the scalar sector contains only a gauge singlet scalar $s$ and the SM Higgs doublet $\phi$. The mass eigenstates at $T=0$ are linear combinations of the Higgs field $h$ and $s$,
\be
H = h\cos\beta + s\sin\beta \,\quad S = -h\sin\beta + s\cos\beta.
\ee
Finally, we also add a fermion field to the singlet sector Lagrangian via
\be
\label{Lpsi}
\mathcal{L}_{\psi} = \bar\psi(i\slashed\partial - m_\psi)\psi + i g s \bar{\psi}\psi ,
\ee
where $\psi$ is a Dirac fermion with a scalar coupling to $s$. 

In this type of hidden sector models where the singlet scalar is assumed to drive inflation in the early universe, depending on the parameter values and possible discrete symmetries, either $s$ or $\psi$ can act as a DM candidate as has been investigated in detail e.g. in \cite{Lerner:2009xg,Kahlhoefer:2015jma,Aravind:2015xst}. In the present study we concentrate on the electroweak phase transition only, and take ${\cal L}_{\psi}$ to simulate the effects of additional degrees of freedom and their interactions on the running of couplings affecting the inflationary dynamics. However, we will comment on the general feasibility of these fields to act as dark matter as well.

\section{Cosmic Inflation}
\label{cosmicinflation}

We begin by considering the cosmic inflation. Here we consider only the part of the action which is relevant for gravitational dynamics, and concentrate first on the general setup allowing both the Higgs field $h$ and the singlet $s$ to be non-minimally coupled to gravity. The analysis can be easily generalized to cover also other inflationary models with or without a non-minimal coupling to gravity.

\subsection{Inflationary dynamics}

The action in the Jordan frame is
\be
\label{nonminimal_action}
S_J = \int d^4x \sqrt{-g}\bigg((\partial_{\mu}\phi)^\dagger \partial^{\mu}\phi + \frac{1}{2} \partial_{\mu}s\partial^{\mu}s - \frac{1}{2}M_{\rm P}^2R-\xi_{\rm h}(\phi^{\dagger}\phi)R -f(s)R - V(s,\phi) \bigg),
\ee
where $M_{\rm P}$ is the reduced Planck mass, $R$ is the Ricci scalar, $\xi_{\rm h}$ is a dimensionless coupling constant, and the function $f(s)$ defines the non-minimal coupling of the $s$-field. We work in the unitary gauge so that $\xi_{\rm h}(\phi^{\dagger}\phi)R = \xi_{\rm h}h^2R/2$.

The non-minimal couplings appearing in the Jordan frame action \eqref{nonminimal_action} can be removed by a conformal transformation to the Einstein frame. Explicitly, the transformation is 
\be
\label{Omega}
\tilde{g}_{\mu\nu} = \Omega^2 g_{\mu\nu}, \hspace{1cm} \Omega^2\equiv 1+\frac{2f(s)}{M_{\rm P}^2} + \frac{\xi_{\rm h} h^2}{M_{\rm P}^2}.
\ee
Then, by a field redefinition
\be
\label{h_chi}
\frac{d\chi_{\rm h}}{dh} = \sqrt{\frac{\Omega^2+6\xi^2_{\rm h}h^2/M_{\rm P}^2}{\Omega^4}}, \hspace{1cm} \frac{d\chi_{\rm s}}{ds} = \sqrt{\frac{\Omega^2+6(df/ds)^2/M_{\rm P}^2}{\Omega^4}},
\ee
we obtain
\be
\begin{aligned}
S_E =& \int d^4x \sqrt{-g}\bigg(-\frac{1}{2}M_{\rm P}^2R + \frac{1}{2}{\partial}_{\mu}\chi_{\rm h}{\partial}^{\mu}\chi_{\rm h} + \frac{1}{2}{\partial}_{\mu}\chi_{\rm s}
{\partial}^{\mu}\chi_{\rm s} \\
&+ A(\chi_{\rm s}, \chi_{\rm h}){\partial}_{\mu}\chi_{\rm h}{\partial}^{\mu}\chi_{\rm s} - U(\chi_{\rm s},\chi_{\rm h})  \bigg),
\label{EframeS}
\end{aligned}
\ee
where $U(\chi_{\rm s},\chi_{\rm h}) = \Omega^{-4}V(s(\chi_{\rm s}),h(\chi_{\rm h}))$ and
\be
A(\chi_{\rm s}, \chi_{\rm h}) = \frac{6\xi_{\rm h}(df/ds)}{M_{\rm P}^2\Omega^4}\frac{ds}{d\chi_{\rm s}}\frac{dh}{d\chi_{\rm h}}h.
\ee
Note that in writing Eq. (\ref{EframeS}) we have dropped the tilde as everything is written in terms of the transformed coordinates. 

This general setup allows many realizations of inflationary dynamics. In the following we will consider the $s$-inflation scenario \cite{Lerner:2009xg}, i.e. inflation occurring in the $s$-direction. We will therefore take 
$f(s)=\xi_{\rm s}s^2/2$ and set the Higgs field vacuum expectation value to $h=0$. Restricting to this simple non-minimal coupling is motivated also by the analysis of quantum corrections in a curved background which have been shown to generate a term of this form even if $\xi_s$ is initially set to zero \cite{Birrell:1982ix}. For discussion concerning the effect of other gravitational couplings, such as $\alpha R^2$, see e.g. \cite{Salvio:2015kka,Calmet:2016fsr}.

Consistency of the $s$-inflation scenario requires that the minimum of the potential at large $s$ and $h$ is very close to the $h=0$ direction. This is true if $\lambda_{\rm s}/\xi^2_{\rm s}\ll \lambda_{\rm h}/\xi^2_{\rm h}$. Consequently, $A(\chi_{\rm s}, \chi_{\rm h})=0$ and the kinetic terms of the scalar fields are canonical. Furthermore, we assume hierarchy between the non-minimal couplings of the scalar fields to gravity, $\xi_{\rm h}\ll \xi_{\rm s}$.

When considering inflation, in the Jordan frame we can focus on the direction along the inflaton field and need to take into account only the highest order term in the potential. Hence, the relevant part of the potential is $V(s,\phi)=\lambda_{\rm s} s^4/4$. In the Einstein frame the potential becomes at large field values
\be
\label{chipotential}
U(\chi_{\rm s}) \simeq \frac{\lambda_{\rm s} M_P^4}{4\xi_{\rm s}^2}\left(1+\exp\left(-\frac{2\sqrt{\xi_{\rm s}}\chi_{\rm s}}{\sqrt{6\xi_{\rm s}+1}M_P} \right) \right)^{-2} .
\ee
At high field values, $\chi_{\rm s}\gg M_{\rm P}$ or equivalently at 
$s\gg M_{\rm P}/ \xi_{\rm s}^{1/2}$, this is a sufficiently flat potential to drive inflation.\footnote{If inflation occurs along the $h$-direction, i.e. if $\lambda_{\rm s}/\xi^2_{\rm s}\gg \lambda_{\rm h}/\xi^2_{\rm h}$, an analogous expression is obtained. This is the case corresponding to the Higgs inflation \cite{Bezrukov:2007ep}.}

The inflationary dynamics is characterized by the usual slow-roll parameters and total number of e-folds during inflation. The slow-roll parameters are defined in terms of the Einstein frame potential by
\bea
\epsilon &\equiv& \frac{1}{2}M_P^2 \left(\frac{1}{U}\frac{{\rm d}U}{{\rm d}\chi_{\rm s}}\right)^2 , \\ \nonumber
\eta &\equiv& M_P^2 \frac{1}{U}\frac{{\rm d}^2U}{{\rm d}\chi_{\rm s}^2} ,
\eea
and the number of e-folds by
\be
N = \frac{1}{M_P^2} \int_{s_f}^{s_i} {\rm d}s \left(\frac{{\rm d}\chi_{\rm s}}{{\rm d} s}\right)^2 U \left(\frac{{\rm d}U}{{\rm d} s}\right)^{-1},
\label{Ndef}
\ee
where the field value at the end of inflation, $s_f$, is defined via $\epsilon(s_f) = 1$, and for a given value of $N$ Eq. (\ref{Ndef})  determines the field value $s_i$ at the time the corresponding scales left the horizon. 

To obtain the correct amplitude for the curvature power spectrum the COBE normalization requires \cite{Lyth:1998xn}
\be
\label{cobe}
\frac{U(s_i)}{\epsilon(s_i)} = (0.027M_P)^4 ,
\ee
which at tree-level can be expressed in terms of the required e-folds and $s$-field couplings as
\be
\label{xi}
\frac{2\lambda_{\rm s} N^2}{\xi_{\rm s}(6\xi_{\rm s}+1)}=0.027^4.
\ee
This equation, at tree-level, determines the required value of the non-minimal coupling $\xi_{\rm s}$ in terms of $\lambda_{\rm s}$ and $N$. For instance, 
$\lambda_{\rm s}\simeq 0.1$ requires $\xi_{\rm s}=\mathcal{O}(10^4)$ for $N=60$. 
Despite the claims on possible unitarity violation during inflation \cite{Burgess:2009ea,Barbon:2009ya,Barvinsky:2009ii,Bezrukov:2010jz,Burgess:2010zq,Hertzberg:2010dc}, this large non-minimal coupling to gravity is not necessarily problematic for the SM Higgs inflation \cite{Calmet:2013hia}; see also \cite{Burgess:2014lza,Bezrukov:2014ipa, Fumagalli:2016lls, Enckell:2016xse} for more recent discussion.
On the other hand, for $s$-inflation this problem does not arise at all
because the scale of perturbative unitarity breaking is always higher than the scale of inflation \cite{Calmet:2013hia,Kahlhoefer:2015jma}.

For the spectral index, $n_s-1 \simeq -6\epsilon+2\eta$, and tensor-to-scalar ratio, $r\simeq 16\epsilon$, we obtain the following tree-level estimates
\bea
\label{nsr}
n_s^{\rm tree}(s_i) &\simeq& 1 -\frac{2}{N} - \frac{3(6\xi_{\rm s}+1)}{4\xi_{\rm s} N^2} , \nn \\
r^{\rm tree}(s_i) &\simeq& \frac{2(6\xi_{\rm s}+1)}{\xi_{\rm s} N^2} .
\eea
For $N=60$ and $\xi_{\rm s}\gg 1$, the numerical values are $n_s^{\rm tree}=0.965$, $r^{\rm tree}=3\times10^{-3}$, in accord with the Planck results $n_s=0.9677\pm 0.0060$ ($68\%$ confidence level) and $r<0.11$ ($95\%$ confidence level) \cite{Ade:2015lrj}.

However, the above expressions hold only at tree-level. To take into account the effect of quantum corrections to the effective potential, we compute the renormalization group evolution of different couplings in the Jordan frame and determine the inflationary observables numerically from the Einstein frame potential $U(\chi_{\rm s},\chi_{\rm h}) = \Omega^{-4}V(s(\chi_{\rm s}),h(\chi_{\rm h}))$, where $\Omega$ is given by \eqref{Omega} and $s(\chi_{\rm s}),h(\chi_{\rm h})$ by \eqref{h_chi}. The beta functions are given in Appendix \ref{betafunctions}. The strength of the non-minimal coupling $\xi_{\rm s}(s_i)$ is fixed by \eqref{cobe} and we use $N=60$. Uncertainty in $N$ corresponds to an uncertainty $\Delta n_s \simeq 2\Delta N/N^2$ in the spectral index. We will show, in sec. \ref{reheating}, that for $s$-inflation $58<N<61$, so $\Delta n_s\lsim0.0011$. As this uncertainty is typically smaller than the effect of loop corrections, using a constant $N$ provides a reasonable estimate for the deviation from the tree-level predictions.

\begin{figure}
\begin{center}
\includegraphics[height=0.46\textwidth]{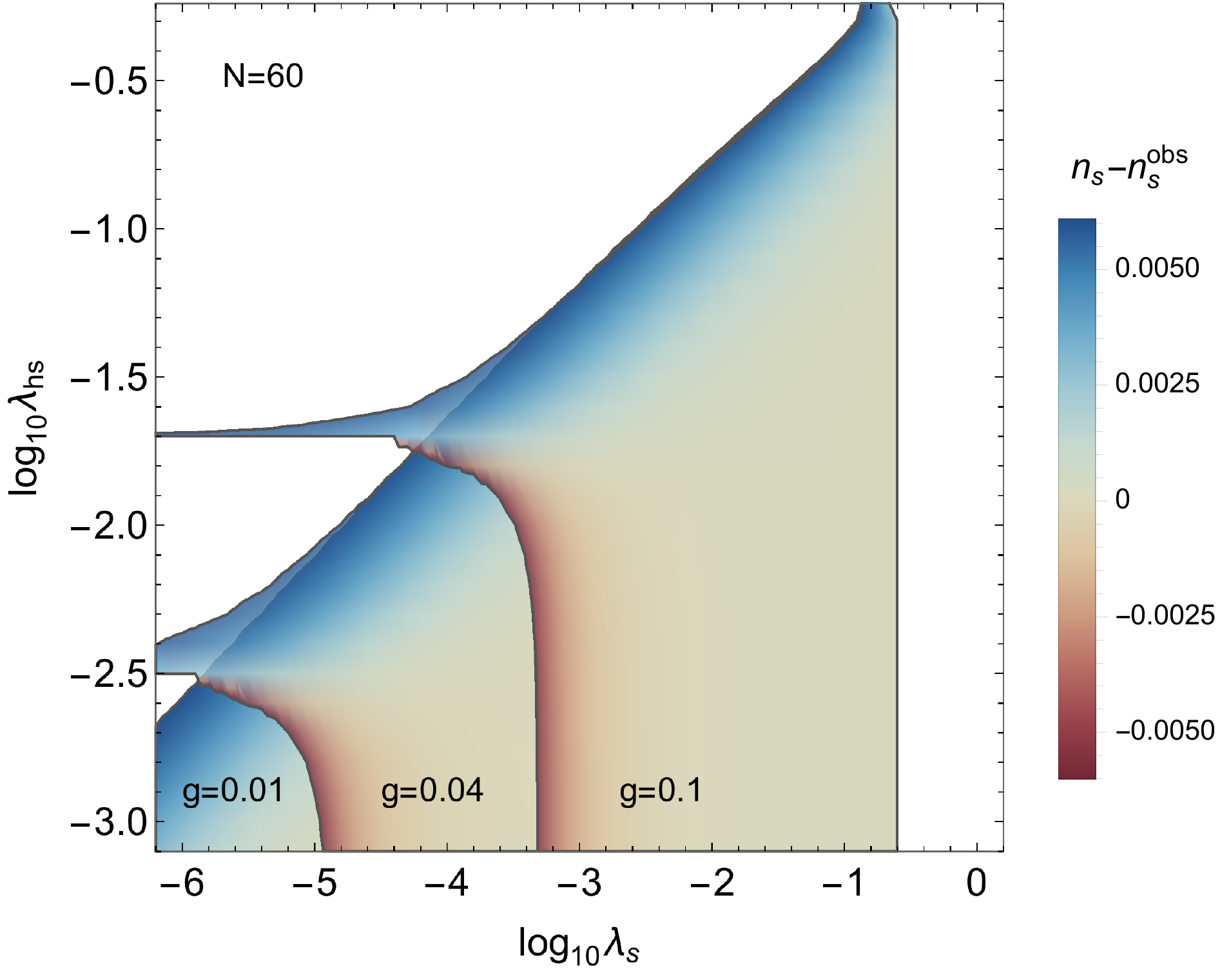}
\caption{Colored regions show where the spectral index satisfies the $1\sigma$ Planck bound $n_s=0.9677\pm 0.0060$ for three different values of $g$. The tensor-to-scalar ratio satisfies the Planck $2\sigma$ bound $r<0.11$ in every point. All coupling values are given at the scale $m_{\rm t}$.}
\label{inflation}
\end{center}
\end{figure}

In Figure \ref{inflation}, the colored regions show the values of the couplings $\lambda_{\rm s}(m_{\rm t})$ and $\lambda_{\rm hs}(m_{\rm t})$, which give $n_s=0.9677\pm 0.0060$. The three different regions correspond to values $g=0.1$, $0.04$, and $g=0.01$. When scanning the parameter space, we have required all scalar couplings to remain free of Landau poles under the one-loop evolution up to the inflationary scales, $s\gsim M_{\rm P}/ \xi_{\rm s}^{1/2}$. This constraint sets the rightmost boundary of the shaded areas in Figure \ref{inflation}.\footnote{In the numerical computation we have used the cutoff value $\lambda_{\rm s}\le 4\pi$ at all scales considered. The exact location of this boundary is, however, only very mildly dependent on the value of the exact value of this upper bound, since $\lambda_{\rm s}$ evolves very rapidly to infinitely large values after it becomes strong.} In the white region above and to the left from the colored contours, $n_s$ is larger than $0.9677+0.0060$. In the white region towards the lower left corner of the plot, $n_s$ becomes smaller than $0.9677-0.0060$. The tensor-to-scalar ratio satisfies $r<0.11$ in every point.

\subsection{Reheating in $s$-inflation}
\label{reheating}

We need to ensure that reheating in this model occurs at sufficiently high temperature, $T_{\rm{RH}}\gg T_{\rm{EW}}$, and that the obtained number of inflationary e-folds is consistent with the above analysis. Reheating in $s$-inflation has been discussed extensively in \cite{Lerner:2011ge} (see also \cite{Bezrukov:2008ut}). 

In $s$-inflation reheating occurs by production of $s$ and $h$ particles, which soon annihilate or decay to other SM particles. For sufficiently large $\lambda_{\rm s}$, the possible decay channels for an oscillating $s$ condensate are decay by stochastic resonance to Higgs bosons and production of inflaton excitations, i.e. $s$ particles, in the quadratic part of the $s$ potential. The first channel requires $\lambda_{\rm s} > 0.25\lambda_{\rm hs}$ and the latter $\lambda_{\rm s}>0.019$ \cite{Lerner:2011ge}. However, because we want to study field dynamics also in the region of the parameter space where the scalar self-interaction is very small, $\lambda_{\rm s}(m_{\rm t})\gtrsim 10^{-7}$, we have to extend the results obtained in \cite{Lerner:2011ge}.

Reheating can occur not only in the quadratic but also in the quartic part of the $s$ potential. Transition into quartic potential occurs at $s\simeq M_{\rm P}/\xi_{\rm s}$ \cite{Bezrukov:2008ut}, and in the following we assume $10^{-9}<\lambda_{\rm s} < 10^{-2}$ at the time of reheating. In the quartic potential the homogeneous $s$ field evolves as $s(t)=\sigma_0(t){\rm cn}(0.85\lambda_{\rm s}^{1/2}\sigma_0(t) t, 1/\sqrt{2})$, where cn is the Jacobi cosine and $\sigma_0$ a time-dependent oscillation amplitude. For the singlet condensate $s_0$ decay rate into $s$ field quanta we use \cite{Nurmi:2015ema,Kainulainen:2016vzv}
\be
\Gamma_{s_0\rightarrow ss}^{(4)} = 0.023\lambda_{\rm s}^{3/2}\sigma_0 .
\ee
Decay to Higgs bosons is assumed to be inefficient due to large, rapidly forming thermal masses.

During reheating the $s$ field governs the evolution of the universe, and in the quartic potential the Hubble parameter is given by $3H^2M_{\rm P}^2=\lambda_{\rm s}/4\sigma_0^4$. The condensate decay by $s_0\rightarrow ss$ becomes efficient when
\be
\frac{\Gamma_{s_0\rightarrow ss}^{(4)}}{H} = 0.08\lambda_{\rm s}\frac{M_{\rm P}}{\sigma_0} \simeq 1 ,
\ee
giving $\sigma_0 \gtrsim 1.9\times 10^{8}$ GeV for $\lambda_{\rm s}(\sigma_0)\gtrsim 10^{-9}$. For sufficiently large portal coupling values the corresponding energy density, $\rho_{s_0}=\lambda_{\rm s}/4\sigma_0^4$, is rapidly converted to SM particles by $s$ annihilations and decays. Equating this with the energy density $\pi^2/30g_* T^4$ of the heat bath consisting of the SM and $s$ fields, we obtain for the reheating temperature
\be
\label{RHtemp}
T_{\rm RH} \gtrsim 0.02\lambda_{\rm s}^{5/4} M_{\rm P} \gtrsim 3.1\times 10^5  {\rm GeV},
\ee
where we used $g_* = 107.75$ and $\lambda_{\rm s}\gtrsim 10^{-9}$. The result \eqref{RHtemp} shows that $T_{\rm RH}\gg T_{\rm EW}$ for all parameter values used, which means that the fields have in all cases relaxed to their minima at the time the electroweak phase transition occurs.

The number of inflationary e-folds is then given by
\be
\label{efolds}
N = \ln\left(\left(\frac{\rho_{\rm RD}}{\rho_{\rm end}}\right)^{1/3}\left(\frac{\rho_{\rm RH}}{\rho_{\rm RD}}\right)^{1/4}\left(\frac{g_0T_0^3}{g_*T_{\rm RH}^3} \right)^{1/3}H_k\lambda \right) \simeq 60-\frac{1}{12}\ln\left(\lambda_{\rm s}N^4 \right),
\ee
where $\rho_{\rm RH}=\pi^2/30g_*T_{\rm RH}^4$, $\rho_{\rm end}\simeq H_k^2M_{\rm P}^2$, $g_0(T_0)\simeq 2$, and $T_0=2.725 K$ is the present-day photon temperature. For the Hubble parameter at the WMAP pivot scale, $\lambda\simeq 2\pi/0.002$ Mpc, we use $H_k=8\times 10^{13}\sqrt{r/0.1}{\rm GeV}\simeq 1.4\times 10^{13}$ GeV. For the energy density at the time the scalar undergoes a transition to the quartic potential and the universe enters into a radiation dominated era, we use $\rho_{\rm RD}=\lambda_{\rm s}M_{\rm P}^4/(9\xi_{\rm s}^4)$ \cite{Bezrukov:2008ut}, where $\lambda_{\rm s}/\xi_{\rm s}^2$ is set by \eqref{xi}. 

The result \eqref{efolds} gives $59<N<60$ for $10^{-9}<\lambda_{\rm s} < 10^{-2}$. Note that the smaller the value of $\lambda_{\rm s}$ is, the larger is the required number of e-folds. This is because small $\lambda_{\rm s}$ requires small $\xi_{\rm s}$ in order to produce the measured curvature perturbation, \eqref{xi}, and the smaller the value of $\xi_{\rm s}$ is, the earlier the transition into a radiation dominated universe occurs. As a result, the number of e-folds does not depend on $T_{\rm RH}$.

As shown in \cite{Lerner:2011ge}, a similar result for the number of e-folds holds also for $\lambda_{\rm s} > 10^{-2}$. Adding an additional theoretical uncertainty $N=\pm 1$ to both results gives $58<N<61$, corresponding to $\Delta n_s \simeq 2\Delta N/N^2\simeq 0.0011$ for the maximum theoretical error in the spectral index when using $N=60$. For sufficiently large couplings, $\lambda_{\rm hs}, \lambda_{\rm s}=\mathcal{O}(0.1)$, the error is smaller, $\Delta n_s \ll 0.001$.

\section{Electroweak phase transition}
\label{EWPT}

After the cosmic inflation and reheating of the universe the standard hot big bang scenario, where evolution of the universe is governed by a heath bath consisting of different particle species, is recovered. We now demonstrate how to obtain a strong first order electroweak phase transition at $T=T_{\rm EW}\lsim 150$ GeV with the singlet scalar $s$.  

In the model under consideration, a strong electroweak phase transition can be realized already at tree-level if the transition happens from a minimum in $s$-direction to the electroweak broken minimum at $(h,s)=(v,0)$. Similarly as in \cite{Kainulainen:2015sva}, we define the $s$ field such that its vacuum expectation value at the electroweak broken minimum is $\langle s \rangle =0$. This sets $\mu_{\rm h}^2 = -v^2\lambda_{\rm h}$ and $\mu_1^3 = -v^2\mu_{\rm hs}/2$. To study the phase transition, we include the leading finite-temperature corrections to the scalar potential as
\be
\mu_1(T)^3=\mu_1^3 + c_1 T^2,\quad \mu_{\rm s}(T)^2=\mu_{\rm s}^2+c_{\rm s} T^2,\quad \mu_{\rm h}(T)^2=\mu_{\rm h}^2+c_{\rm h} T^2,
\ee
where
\be
\begin{aligned}
c_1 &= \frac{1}{12}(\mu_3 + 2\mu_{\rm hs} + 2gm_\psi), \\
c_{\rm s} &= \frac{1}{12}(2\lambda_{\rm hs}+3\lambda_{\rm s} + 2g^2), \\
c_{\rm h} &= \frac{1}{48}(9 g_L^2+3g_Y^2+12 y_t^2+24\lambda_{\rm h}+2\lambda_{\rm hs}).
\end{aligned}
\ee
Here $g_L$ and $g_Y$ are the weak and hypercharge gauge couplings and $y_t$ is the top Yukawa coupling. Next, we will first study a $Z_2$ symmetric potential for which analytical results can be derived, and then consider general potential \eqref{scalarpotential}.

\subsection{$Z_2$-symmetric potential}
Requiring $Z_2$ symmetry for the scalar potential sets $\mu_1 = \mu_3 = \mu_{\rm hs} = 0$ in Eq. (\ref{scalarpotential}). For the $Z_2$ symmetric potential the conditions for a strong first order electroweak phase transition can be derived analytically. First, a nontrivial minimum must exist along the $s$-direction, requiring $\mu_{\rm s}^2 < 0$. Second, the electroweak breaking minimum along the $h$-direction at $h=v=246{\rm GeV}$ must be the deepest minimum at $T=0$. This is realized if 
\be
\label{cond1}
\frac{\mu_{\rm s}^4}{\lambda_{\rm s}} < \frac{\mu_{\rm h}^4}{\lambda_{\rm h}}.
\ee 
Third, in order to get the required phase transition pattern, the minimum in the $s$-direction must appear at higher temperature than the one along the $h$-direction. The condition for this is
\be 
\label{cond2}
\frac{\mu_{\rm s}^4}{c_{\rm s}^2} > \frac{\mu_{\rm h}^4}{c_{\rm h}^2}.
\ee 
The mass of $s$ is $m_{\rm s}^2 = \mu_{\rm s}^2 + \lambda_{\rm hs}v^2/2$, so the condition \eqref{cond2} can be written as a lower limit on the portal coupling. On the other hand, \eqref{cond1} provides an upper limit for the portal coupling, so that
\be
\label{ewptcond1}
\frac{2m_{\rm s}^2}{v^2} + \frac{2 \lambda_{\rm h} c_{\rm s}}{c_{\rm h}} < \lambda_{\rm hs} < 2\sqrt{\lambda_{\rm h}\lambda_{\rm s}} + \frac{2m_{\rm s}^2}{v^2}.
\ee
Here we have used the relation $\mu_{\rm h}^2 = -\lambda_{\rm h} v^2$. 

The critical temperature $T_c$ at which the two extrema are degenerate is given by
\be
\label{criticalT}
T_c^2 = \frac{c_{\rm s} \mu_{\rm s}^2 \lambda_{\rm h} + \sqrt{\lambda_{\rm h}\lambda_{\rm s} c_{\rm h}^2(\mu_{\rm s}^4+\lambda_{\rm h}\lambda_{\rm s}v^4) - c_{\rm s}^2\lambda_{\rm h}^2v^4}}{c_{\rm h}^2\lambda_{\rm s} - c_{\rm s}^2\lambda_{\rm h}}.
\ee
To realize a first order electroweak phase transition we must also require that the extremum in the $s$-direction is a minimum.  At $T_c$, the condition is 
\be
\label{ewptcond2}
\lambda_{\rm hs} > 2\sqrt{\lambda_{\rm h}\lambda_{\rm s}},
\ee 
and below $T_c$ a somewhat more constraining 
\be
\lambda_{\rm hs} (\mu_{\rm s}^2 + c_{\rm s} T^2) < 2 \lambda_{\rm s} (-\lambda_{\rm h} v^2 + c_{\rm h} T^2).
\ee 
Finally, for a successful electroweak baryogenesis the transition has to be strong,
characterized by the requirement $v(T_c)/T_c>1$. Here
\be
v(T)^2 = v^2 - \frac{c_{\rm h} T^2}{\lambda_{\rm h}},
\ee
is the vacuum expectation value of $h$ at temperature $T$.

\begin{figure}
\begin{center}
\includegraphics[height=0.46\textwidth]{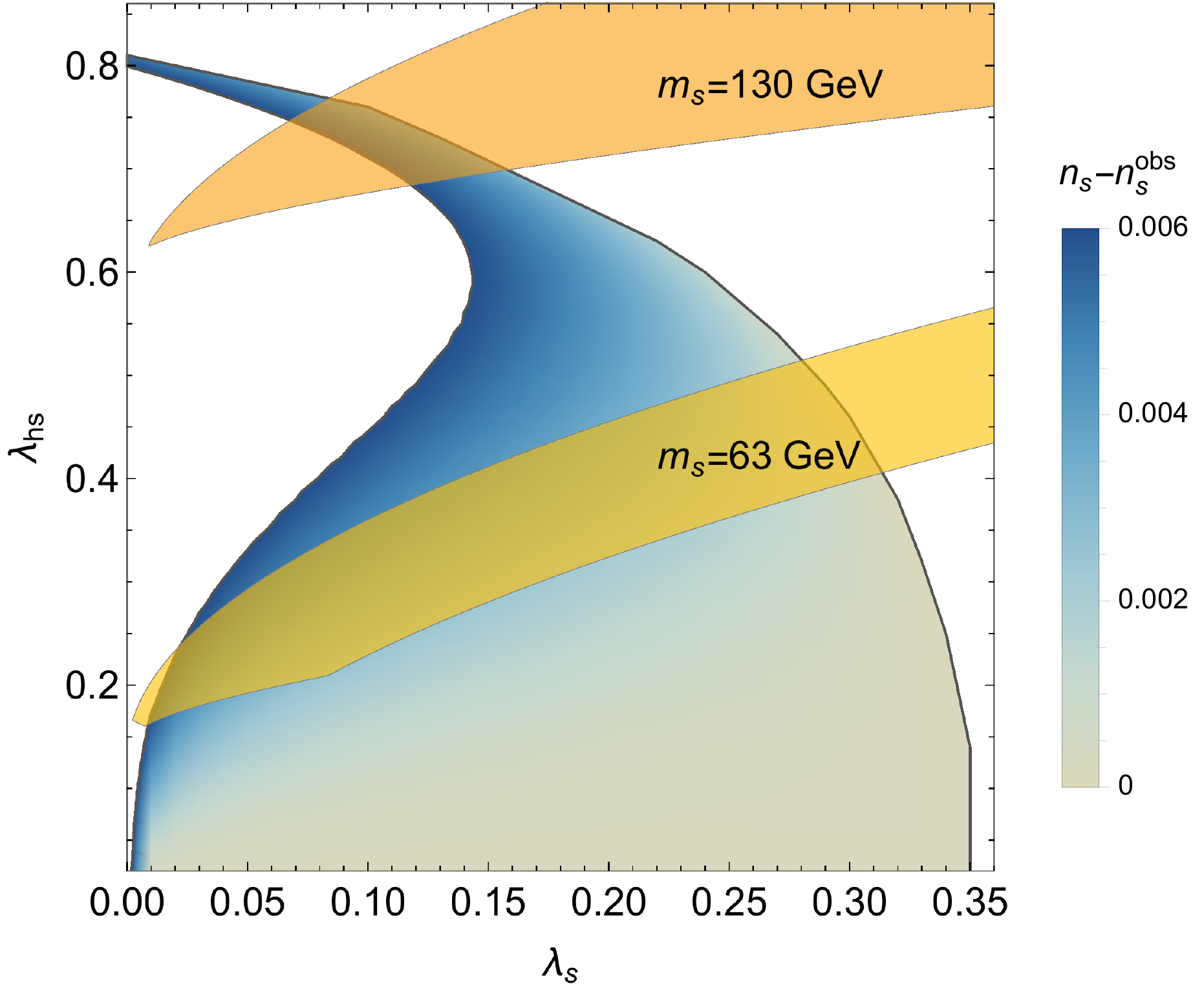}
\caption{The yellow and orange regions show, for different scalar masses, where a $Z_2$-symmetric $s$ gives a strong electroweak phase transition. The shaded region shows where the spectral index satisfies the $1\sigma$ Planck bound $n_s=0.9677\pm 0.0060$. Here $g=0$.}
\label{inflation+ewpt_Z2only}
\end{center}
\end{figure}

If $s$ is stable, its mass and coupling to the Higgs are heavily constrained by the dark matter overclosure constraint and constraints from direct dark matter searches. As shown in \cite{Cline:2013gha}, only the region near the Higgs resonance, $m_{\rm s}\sim m_{\rm h}/2$, and the region $m_{\rm s}\gsim m_{\rm h}$ are allowed. When the masses are increased in the high mass region, $m_{\rm s}\gsim m_{\rm h}$, the portal coupling of $\lambda_{\rm hs}={\cal O}(1)$ is required to obtain a strong first order electroweak phase transition. As shown in Figure \ref{inflation+ewpt_Z2only}, this large value of $\lambda_{\rm hs}$ is in tension with successful $s$-inflation. Therefore, for our purposes only the region $m_{\rm s}\simeq m_{\rm h}$ is interesting. As further shown in Figure \ref{inflation+ewpt_Z2only}, both strong electroweak phase transition and $s$-inflation can be successfully realized in this region.

\subsection{General potential}
Next, we consider the full scalar potential \eqref{scalarpotential}, and include also a singlet fermion in the hidden sector. 
We perform a Monte Carlo scan over the following parameter space:
\be
\begin{gathered}
0.5<m_{\rm S}/{\rm GeV}<126\,, \quad m_\psi =4m_{\rm S}\,, \\ \quad 10^{-6}<|\mu_{\rm hs}/{\rm GeV}|<10^2\,, \quad 10^{-6}<|\mu_3/{\rm GeV}|<10^2\,, \\ 10^{-7}<\lambda_s<0\,, \quad 10^{-4}<\lambda_{\rm hs}<0\,, \quad 0.02<g<5\,.
\end{gathered}
\ee
We first check the Higgs boson couplings against the LHC data \cite{Aaltonen:2013ioz,Khachatryan:2014jba,Aad:2015gba} to constrain the Higgs boson mixing angle $\cos\beta$ and its invisible decay width. Then, for each parameter set we check, by analyzing the potential numerically, if they give a strong first order electroweak phase transition, $v(T_c)/T_c>1$. For the points which do, we compute the $\psi$ relic density 
\be
f_{\rm rel} = \Omega_\psi h^2/0.12 ,
\ee 
by a standard freeze-out calculation similarly to \cite{Alanne:2014bra}. The dominant annihilation channel is $\psi\bar\psi\to SS$, and we take into account only scalar final states. Further, following \cite{Alanne:2014bra}, we also check the LUX upper limit on spin independent dark matter scattering off nuclei \cite{Akerib:2015rjg}. Finally, we calculate the inflationary observables as described in Section \ref{cosmicinflation}.

\begin{figure}
\begin{center}
\includegraphics[height=0.38\textwidth]{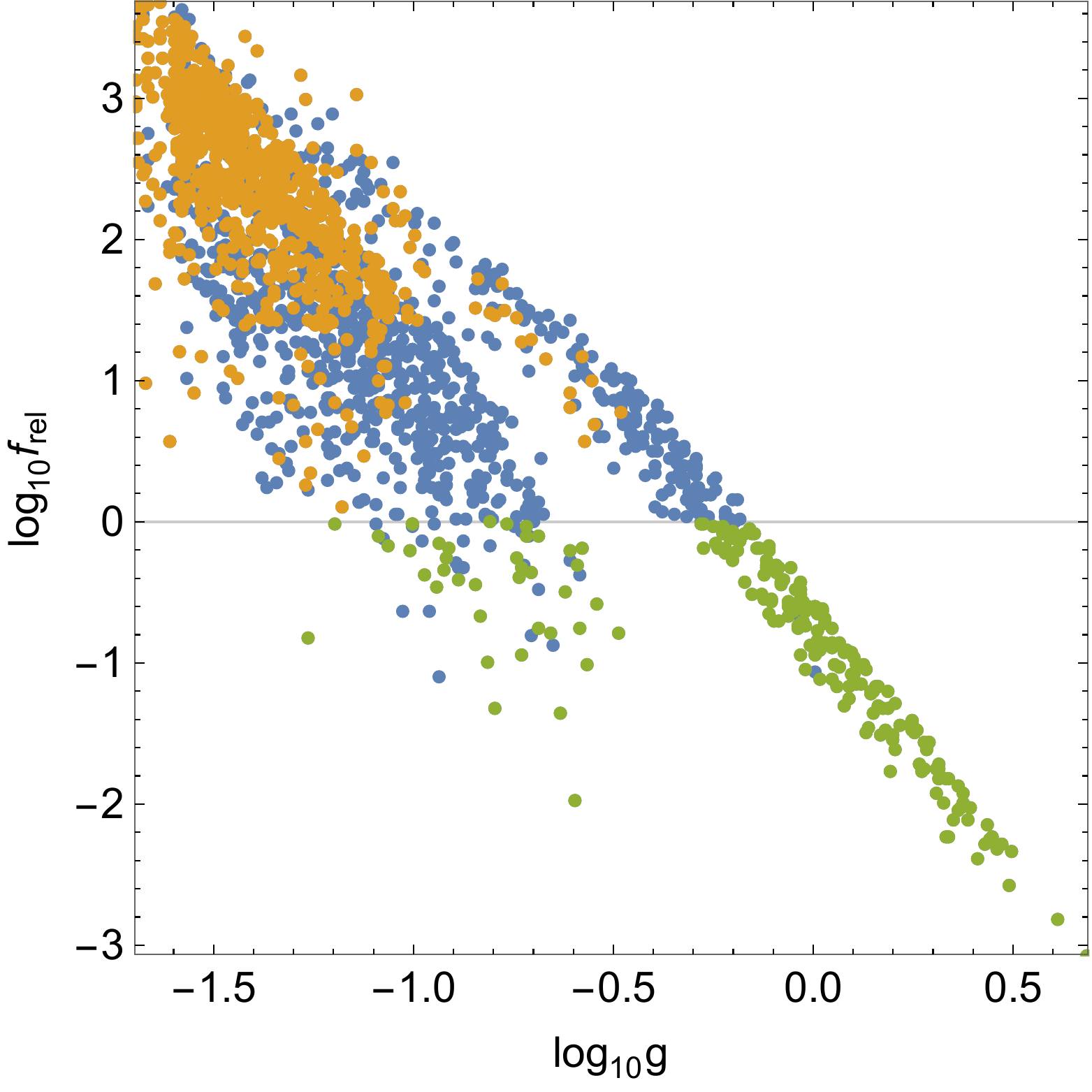} \hspace{0.2cm}
\includegraphics[height=0.38\textwidth]{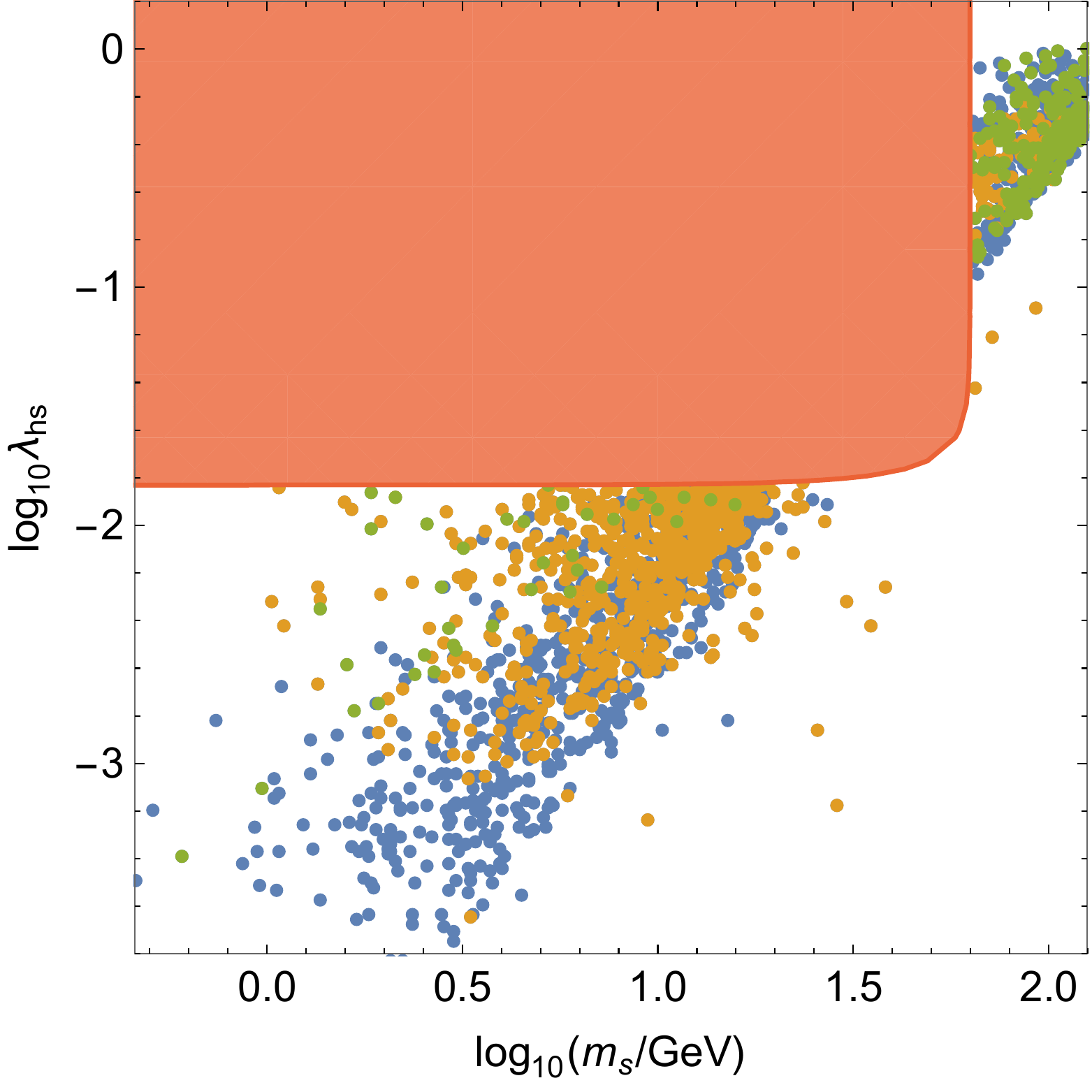} \\ \vspace{0.1cm}
\includegraphics[height=0.38\textwidth]{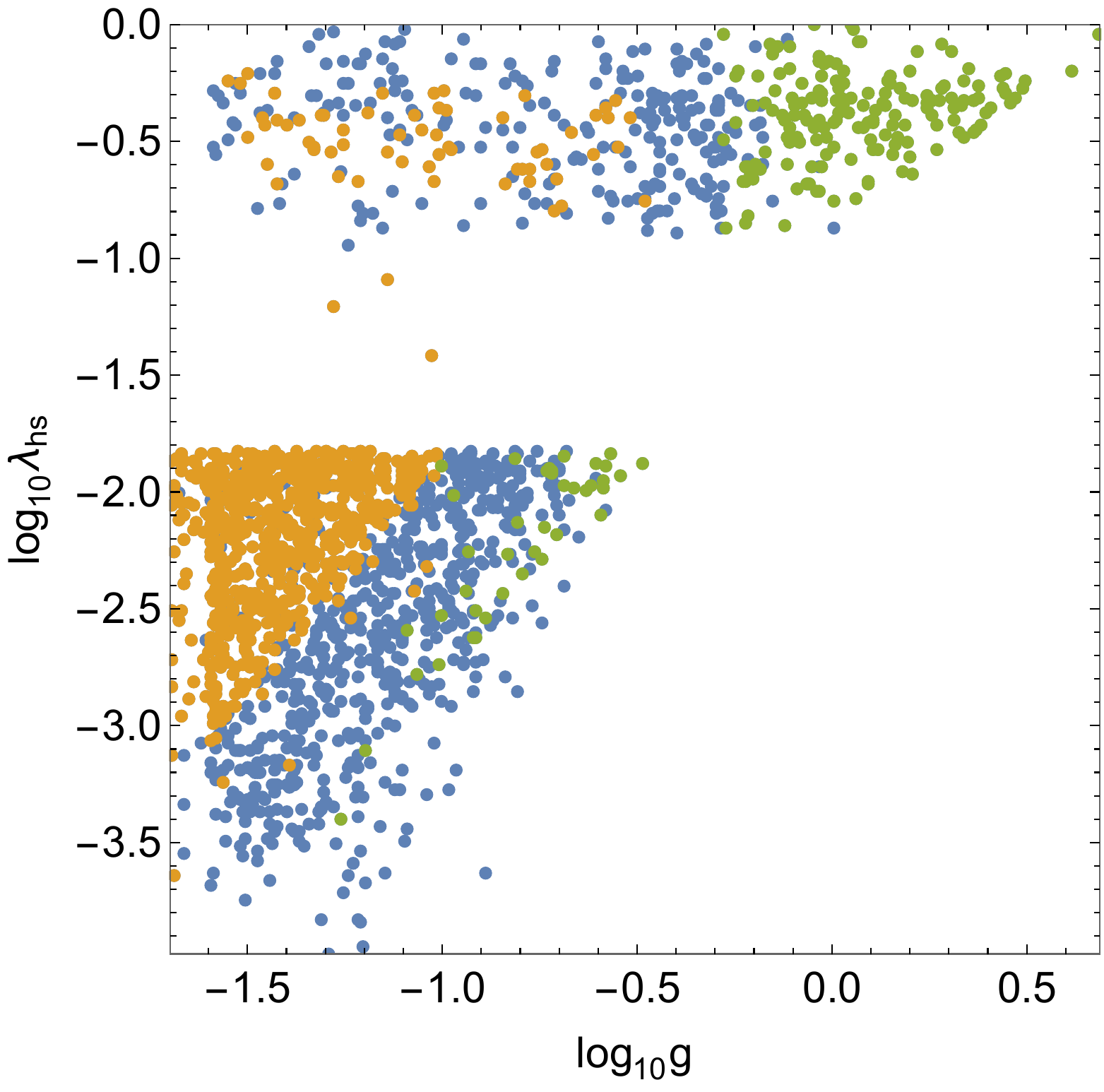} \hspace{0.2cm}
\includegraphics[height=0.38\textwidth]{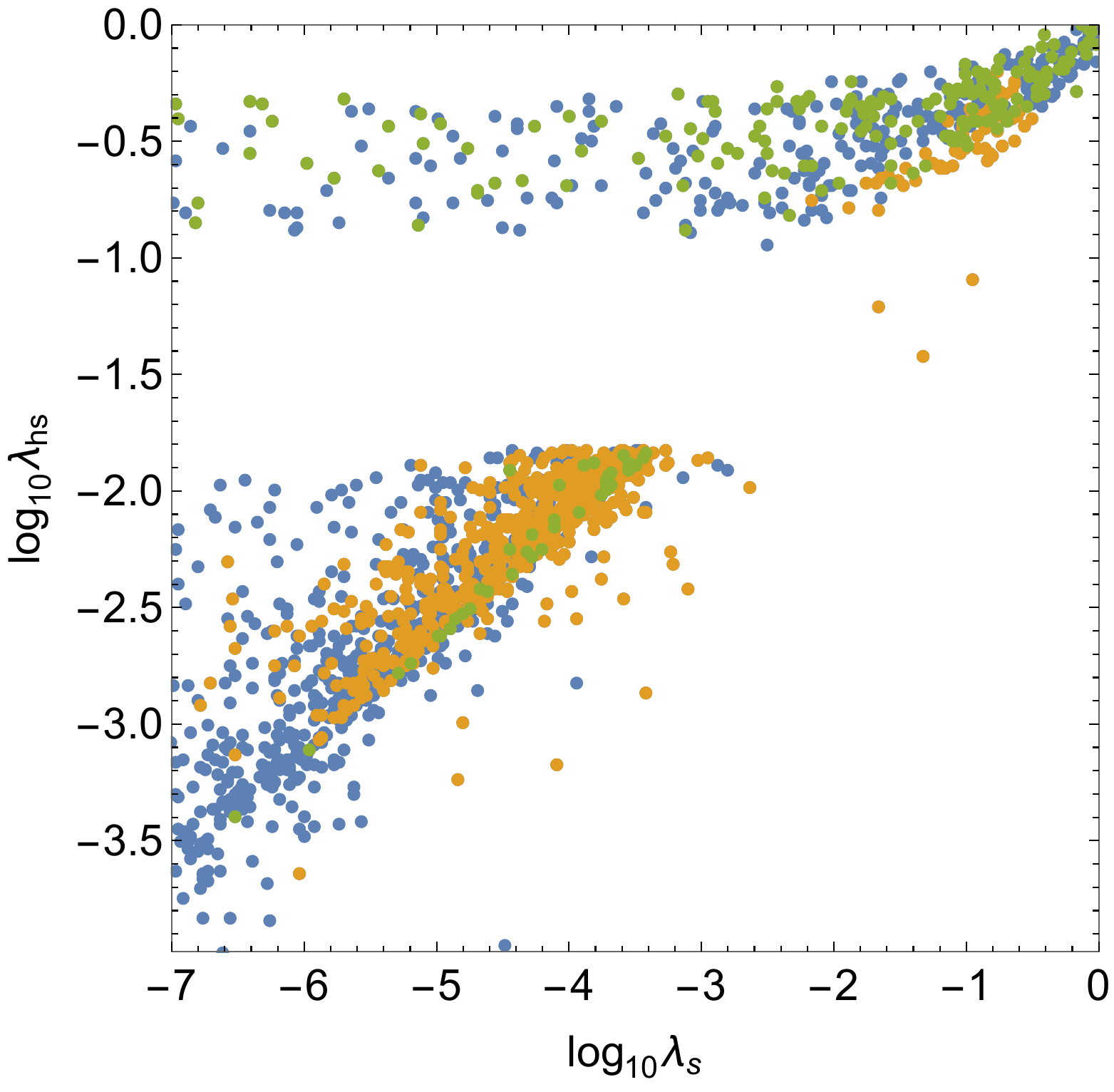} 
\caption{All points give a strong electroweak phase transition and are compatible with the LHC data. For yellow points also inflation is successful. The green points give $f_{\rm rel} <1$ and are in agreement with the LUX constraint. The green and yellow points do not overlap. The red region in the upper right panel is excluded by the LHC constraint on Higgs invisible decay width. In this figure, there is no mixing between sterile and active neutrinos.}
\label{inflation+ewpt+dm}
\end{center}
\end{figure}

In Figure \ref{inflation+ewpt+dm} we show the points compatible with the LHC data and giving a strong first order electroweak phase transition. The gap in the region $-1.8<\log_{10}\lambda_{\rm hs}<-1$ is due to the LHC constraint on Higgs invisible decay width, as shown in the upper right panel. The yellow points are compatible with the constraints due to inflation and the green points both give a sufficiently small dark matter relic density, $f_{\rm rel} <1$, and are compatible with the LUX constraint. The blue points below the $f_{\rm rel}=1$ line in the upper left panel are excluded by LUX.

The plots illustrate a severe tension between obtaining a small enough dark matter relic density and compatibility with the inflationary constraints: the former requires large enough Yukawa coupling $g$ in the singlet sector, while for the latter this same coupling needs to be sufficiently small. As shown in Figure \ref{inflation}, the larger the value of $g$ is the larger also $\lambda_{\rm s}$ has to be for successful inflation. For a strong first order electroweak phase transition also $\lambda_{\rm hs}$ has to be large, as shown in the lower right panel of Figure \ref{inflation+ewpt+dm}, and the larger the value of $\lambda_{\rm hs}$ is the heavier $S$ has to be, as shown in the upper right panel. For this set of points we have fixed the mass hierarchy\footnote{We have experimented with different mass patterns of the singlet scalar and fermion, but the results remain qualitatively similar.} to $m_\psi=4m_{\rm S}$, which means that increasing $\lambda_{\rm hs}$ increases also $m_\psi$. Finally, to get a sufficiently small $\psi$ relic abundance, we also have to increase $g$ if we increase $\lambda_{\rm hs}$. As shown in the lower left panel, it is therefore difficult to reconcile successful inflation and modifications to the electroweak phase transition without, at the same time, overclosing the universe with the singlet fermion which in this simple model setup is a stable particle. 

However, this difficulty is easily avoided. The singlet fermion is essentially a sterile neutrino and we can allow it to mix with an active neutrino $\nu$ with the mixing $\sin\theta$ \cite{Kainulainen:2015sva}. The invisible decay width of the $Z$ boson \cite{Agashe:2014kda}
\be
\frac{\Gamma(Z\rightarrow{\rm{inv}})}{\Gamma(Z\rightarrow \nu\nu)}=2.990\pm 0.007,
\ee
implies that $\sin^2\theta<0.007$ if the contribution from the sterile neutrino decay channel is below $1\sigma$. The mixing with the active neutrinos opens a decay channel for the singlet fermion with the corresponding width
\be
\Gamma_{\psi\rightarrow 3\nu}=\frac{G_F^2}{192\pi^2}m_\psi^5\sin^2\theta.
\ee
To alleviate the overclosure constraint we require that the singlet fermion decays before the big bang nucleosynthesis, i.e. that its lifetime is below 0.1 sec. This leads to a bound
\be
\sin^2\theta>\left(\frac{10\,{\rm MeV}}{m_\psi}\right)^5.
\ee
In the Monte Carlo scan we considered masses $m_\psi$ from 2 GeV to 504 GeV leading, respectively, to lower bounds $\sin^2\theta>3.1\times10^{-12}$ and $\sin^2\theta>3.1\times10^{-24}$. 
Therefore, mixing with the active neutrinos provides an effective decay channel for the singlet fermion. Allowing the singlet fermion to decay to active neutrinos makes all the yellow points viable in Figure \ref{inflation+ewpt+dm}, showing that both inflation and a strong first order electroweak phase transition can be successfully realized within this model.

\section{Conclusions}
\label{conclusions}

In this work we have studied, for the first time, a scenario where the same field that drives inflation can be responsible for a strong electroweak phase transition at a later stage in the history of the universe. We studied this generic setup in a Higgs portal model where the Standard Model is extended with a singlet scalar coupled non-minimally to gravity and assumed to couple sufficiently strongly with the SM Higgs field in order to provide for a strong first order electroweak phase transition.

We identified the regions of the parameter space where the model describes inflation successfully, is compatible with the LHC data, and yields a strong first order electroweak phase transition. We also included a singlet fermion with scalar coupling to the singlet scalar to probe the sensitivity of the constraints on additional degrees of freedom, as they will eventually be necessary if one wants to extend the model to account for sufficient CP violation relevant for electroweak baryogenesis. Finally, we also commented on dark matter production within this model setup and showed how potential overclosure problems can be easily avoided by allowing the singlet fermion to mix with the SM neutrinos.

The order of the electroweak phase transition is interesting not only from the baryogenesis point of view but also for the possibility to observe gravitational waves originating from a first order phase transition. For example, the space-based eLISA detector will have maximum sensitivity at the frequency range relevant for a first order phase transition at the electroweak scale.

Furthermore, while investigation of inflationary models with a non-minimal coupling to gravity is motivated by the effects of quantum corrections in a curved background, our analysis can be easily generalized to cover also other inflationary models with or without a non-minimal coupling to gravity. At the advent of advanced gravitational wave detector era it would indeed be interesting to further study what information could be extracted from inflationary dynamics or other high energy physics by studying in detail phase transitions in the early universe.

\section*{Acknowledgements}
This work has been supported by the Academy of Finland, grant\# 267842. TT acknowledges financial support from the Research Foundation of the University of Helsinki and VV from the Magnus Ehrnrooth foundation.

\appendix

\section{Beta functions}
\label{betafunctions}

There exists some discrepancy in the literature concerning the beta functions of the non-minimal couplings $\xi_{\rm h}, \xi_{\rm s}$, see e.g \cite{DeSimone:2008ei,Lerner:2011ge, Moss:2015gua}. Our results are not very sensitive to the running of these couplings, and in the numerical calculation we have applied the beta functions given in references \cite{DeSimone:2008ei,Lerner:2011ge}.

The beta functions relevant for $s$-inflation at one loop are given by
\begin{align}
\begin{split}
16\pi^2\beta_{\lambda_{\rm h}} =& -6y_t^4 + \sfrac{3}{8}(2g_L^4 + (g_L^2+g_Y^2)^2) + (12y_t^2 - 9g_L^2 - 3g_Y^2)\lambda_h \\
&+ 24\lambda_{\rm h}^2 + \sfrac{1}{2}c_{\rm s}^2\lambda_{\rm hs}^2 ,
\end{split} \\
16\pi^2\beta_{\lambda_{\rm s}} =& 18c_{\rm s}^2\lambda_{\rm s}^2 + 2\lambda_{\rm hs}^2 + 8g^2\lambda_{\rm s} - 8g^4, \\
\begin{split}
16\pi^2\beta_{\lambda_{\rm hs}} =& 4c_{\rm s}\lambda_{\rm hs}^2 + 12\lambda_{\rm h}\lambda_{\rm hs} + 6c_{\rm s}^2\lambda_{\rm s}\lambda_{\rm hs} - \sfrac{3}{2}(3g_L^2+g_Y^2)\lambda_{\rm hs} \\
&+ 6y_t^2\lambda_{\rm hs} + 4g^2\lambda_{\rm hs} ,
\end{split} \\
16\pi^2\beta_{g} =& 5c_{\rm s}g^3, \\
16\pi^2\beta_{\xi_{\rm h}} =& (\xi_{\rm h} - \sfrac{1}{6}) \left(12\lambda_{\rm h} + 6y_t^2 - \sfrac{3}{2}(3g_L^2+g_Y^2)\right) + (\xi_{\rm s} - \sfrac{1}{6}) c_{\rm s} \lambda_{\rm hs} , \\
16\pi^2\beta_{\xi_{\rm s}} =& (\xi_{\rm h} - \sfrac{1}{6}) 4\lambda_{\rm hs} + (\xi_{\rm s} - \sfrac{1}{6}) 6c_{\rm s} \lambda_{\rm s}, \\
16\pi^2\beta_{g_Y} =& \sfrac{41}{6}g_Y^3 , \\
16\pi^2\beta_{g_L} =& -\sfrac{19}{6} g_L^3 , \\
16\pi^2\beta_{g_S} =& -7g_S^3 , \\
16\pi^2\beta_{y_t} =& (-8g_S^2 - \sfrac{9}{4}g_L^2 - \sfrac{17}{12}g_Y^2)y_t + \sfrac{9}{2} y_t^3 ,
\end{align}
where
\be
c_{\rm s} = \frac{1+\xi_{\rm s} s^2/M_P^2}{1+(6\xi_{\rm s}+1)\xi_{\rm s}s^2/M_P^2},
\ee
and we use the following values for the SM couplings at $m_t = 173.34$ GeV \cite{Buttazzo:2013uya}
\be
\lambda_{\rm h} = 0.12774 \,, \quad g_L = 0.64754 \,, \quad g_Y = 0.35940 \,, \quad g_S = 1.1666 \,, \quad y_t = 0.95113 \,.
\ee
The value of the non-minimal coupling $\xi_{\rm s}$ at the inflationary scale $s\simeq M_{\rm P}/\xi_{\rm s}^{1/2}$ is fixed by \eqref{cobe} and it affects running of the SM couplings only through $c_{\rm s}$. This can be important only at $s\simeq M_{\rm P}/\xi_{\rm s}^{1/2}$, rendering the effect of $\xi_{\rm s}$ on running of the SM couplings negligible at other scales. We have checked that the results are not particularly sensitive to $c_{\rm s}$. The non-minimal coupling between the Higgs and $R$ can be chosen freely as long as $\xi_{\rm h}\ll \xi_{\rm s}$.

\bibliography{SinflationEWPT.bib}

\providecommand{\href}[2]{#2}\begingroup\raggedright\begin{thebibliography}{10}

\bibitem{Bezrukov:2007ep}
F.~L. Bezrukov and M.~Shaposhnikov, {\it {The Standard Model Higgs boson as the
  inflaton}},  {\em Phys. Lett.} {\bf B659} (2008) 703--706,
  [\href{http://arxiv.org/abs/0710.3755}{{\tt arXiv:0710.3755}}].

\bibitem{Burgess:2000yq}
C.~Burgess, M.~Pospelov, and T.~ter Veldhuis, {\it {The Minimal model of
  nonbaryonic dark matter: A Singlet scalar}},  {\em Nucl.Phys.} {\bf B619}
  (2001) 709--728, [\href{http://arxiv.org/abs/hep-ph/0011335}{{\tt
  hep-ph/0011335}}].

\bibitem{Ashoorioon:2009nf}
A.~Ashoorioon and T.~Konstandin, {\it {Strong electroweak phase transitions
  without collider traces}},  {\em JHEP} {\bf 07} (2009) 086,
  [\href{http://arxiv.org/abs/0904.0353}{{\tt arXiv:0904.0353}}].

\bibitem{Lerner:2009xg}
R.~N. Lerner and J.~McDonald, {\it {Gauge singlet scalar as inflaton and
  thermal relic dark matter}},  {\em Phys. Rev.} {\bf D80} (2009) 123507,
  [\href{http://arxiv.org/abs/0909.0520}{{\tt arXiv:0909.0520}}].

\bibitem{Belyaev:2010kp}
A.~Belyaev, M.~T. Frandsen, S.~Sarkar, and F.~Sannino, {\it {Mixed dark matter
  from technicolor}},  {\em Phys.Rev.} {\bf D83} (2011) 015007,
  [\href{http://arxiv.org/abs/1007.4839}{{\tt arXiv:1007.4839}}].

\bibitem{Cline:2012hg}
J.~M. Cline and K.~Kainulainen, {\it {Electroweak baryogenesis and dark matter
  from a singlet Higgs}},  {\em JCAP} {\bf 1301} (2013) 012,
  [\href{http://arxiv.org/abs/1210.4196}{{\tt arXiv:1210.4196}}].

\bibitem{Alanne:2014bra}
T.~Alanne, K.~Tuominen, and V.~Vaskonen, {\it {Strong phase transition, dark
  matter and vacuum stability from simple hidden sectors}},  {\em Nucl. Phys.}
  {\bf B889} (2014) 692--711, [\href{http://arxiv.org/abs/1407.0688}{{\tt
  arXiv:1407.0688}}].

\bibitem{Hambye:2008bq}
T.~Hambye, {\it {Hidden vector dark matter}},  {\em JHEP} {\bf 01} (2009) 028,
  [\href{http://arxiv.org/abs/0811.0172}{{\tt arXiv:0811.0172}}].

\bibitem{DiChiara:2015bua}
S.~Di~Chiara and K.~Tuominen, {\it {A minimal model for ${\rm SU}(N)$ vector
  dark matter}},  \href{http://arxiv.org/abs/1506.03285}{{\tt
  arXiv:1506.03285}}.

\bibitem{Kajantie:1996mn}
K.~Kajantie, M.~Laine, K.~Rummukainen, and M.~E. Shaposhnikov, {\it {Is there a
  hot electroweak phase transition at m(H) larger or equal to m(W)?}},  {\em
  Phys. Rev. Lett.} {\bf 77} (1996) 2887--2890,
  [\href{http://arxiv.org/abs/hep-ph/9605288}{{\tt hep-ph/9605288}}].

\bibitem{Hindmarsh:2015qta}
M.~Hindmarsh, S.~J. Huber, K.~Rummukainen, and D.~J. Weir, {\it {Numerical
  simulations of acoustically generated gravitational waves at a first order
  phase transition}},  {\em Phys. Rev.} {\bf D92} (2015), no.~12 123009,
  [\href{http://arxiv.org/abs/1504.03291}{{\tt arXiv:1504.03291}}].

\bibitem{Abbott:2016blz}
{\bf Virgo, LIGO Scientific} Collaboration, B.~P. Abbott et~al., {\it
  {Observation of Gravitational Waves from a Binary Black Hole Merger}},  {\em
  Phys. Rev. Lett.} {\bf 116} (2016), no.~6 061102,
  [\href{http://arxiv.org/abs/1602.03837}{{\tt arXiv:1602.03837}}].

\bibitem{TheLIGOScientific:2016pea}
{\bf the Virgo, The LIGO Scientific} Collaboration, {\it {Binary Black Hole
  Mergers in the first Advanced LIGO Observing Run}},
  \href{http://arxiv.org/abs/1606.04856}{{\tt arXiv:1606.04856}}.

\bibitem{Seoane:2013qna}
{\bf eLISA} Collaboration, P.~A. Seoane et~al., {\it {The Gravitational
  Universe}},  \href{http://arxiv.org/abs/1305.5720}{{\tt arXiv:1305.5720}}.

\bibitem{Kahlhoefer:2015jma}
F.~Kahlhoefer and J.~McDonald, {\it {WIMP Dark Matter and Unitarity-Conserving
  Inflation via a Gauge Singlet Scalar}},  {\em JCAP} {\bf 1511} (2015), no.~11
  015, [\href{http://arxiv.org/abs/1507.03600}{{\tt arXiv:1507.03600}}].

\bibitem{Aravind:2015xst}
A.~Aravind, M.~Xiao, and J.-H. Yu, {\it {Higgs Portal to Inflation and
  Fermionic Dark Matter}},  \href{http://arxiv.org/abs/1512.09126}{{\tt
  arXiv:1512.09126}}.

\bibitem{Birrell:1982ix}
N.~D. Birrell and P.~C.~W. Davies, {\em {Quantum Fields in Curved Space}}.
\newblock Cambridge Monographs on Mathematical Physics. Cambridge Univ. Press,
  Cambridge, UK, 1984.

\bibitem{Salvio:2015kka}
A.~Salvio and A.~Mazumdar, {\it {Classical and Quantum Initial Conditions for
  Higgs Inflation}},  {\em Phys. Lett.} {\bf B750} (2015) 194--200,
  [\href{http://arxiv.org/abs/1506.07520}{{\tt arXiv:1506.07520}}].

\bibitem{Calmet:2016fsr}
X.~Calmet and I.~Kuntz, {\it {Higgs Starobinsky Inflation}},  {\em Eur. Phys.
  J.} {\bf C76} (2016), no.~5 289, [\href{http://arxiv.org/abs/1605.02236}{{\tt
  arXiv:1605.02236}}].

\bibitem{Lyth:1998xn}
D.~H. Lyth and A.~Riotto, {\it {Particle physics models of inflation and the
  cosmological density perturbation}},  {\em Phys. Rept.} {\bf 314} (1999)
  1--146, [\href{http://arxiv.org/abs/hep-ph/9807278}{{\tt hep-ph/9807278}}].

\bibitem{Burgess:2009ea}
C.~P. Burgess, H.~M. Lee, and M.~Trott, {\it {Power-counting and the Validity
  of the Classical Approximation During Inflation}},  {\em JHEP} {\bf 09}
  (2009) 103, [\href{http://arxiv.org/abs/0902.4465}{{\tt arXiv:0902.4465}}].

\bibitem{Barbon:2009ya}
J.~L.~F. Barbon and J.~R. Espinosa, {\it {On the Naturalness of Higgs
  Inflation}},  {\em Phys. Rev.} {\bf D79} (2009) 081302,
  [\href{http://arxiv.org/abs/0903.0355}{{\tt arXiv:0903.0355}}].

\bibitem{Barvinsky:2009ii}
A.~O. Barvinsky, A.~{\relax Yu}. Kamenshchik, C.~Kiefer, A.~A. Starobinsky, and
  C.~F. Steinwachs, {\it {Higgs boson, renormalization group, and naturalness
  in cosmology}},  {\em Eur. Phys. J.} {\bf C72} (2012) 2219,
  [\href{http://arxiv.org/abs/0910.1041}{{\tt arXiv:0910.1041}}].

\bibitem{Bezrukov:2010jz}
F.~Bezrukov, A.~Magnin, M.~Shaposhnikov, and S.~Sibiryakov, {\it {Higgs
  inflation: consistency and generalisations}},  {\em JHEP} {\bf 01} (2011)
  016, [\href{http://arxiv.org/abs/1008.5157}{{\tt arXiv:1008.5157}}].

\bibitem{Burgess:2010zq}
C.~P. Burgess, H.~M. Lee, and M.~Trott, {\it {Comment on Higgs Inflation and
  Naturalness}},  {\em JHEP} {\bf 07} (2010) 007,
  [\href{http://arxiv.org/abs/1002.2730}{{\tt arXiv:1002.2730}}].

\bibitem{Hertzberg:2010dc}
M.~P. Hertzberg, {\it {On Inflation with Non-minimal Coupling}},  {\em JHEP}
  {\bf 11} (2010) 023, [\href{http://arxiv.org/abs/1002.2995}{{\tt
  arXiv:1002.2995}}].

\bibitem{Calmet:2013hia}
X.~Calmet and R.~Casadio, {\it {Self-healing of unitarity in Higgs inflation}},
   {\em Phys. Lett.} {\bf B734} (2014) 17--20,
  [\href{http://arxiv.org/abs/1310.7410}{{\tt arXiv:1310.7410}}].

\bibitem{Burgess:2014lza}
C.~P. Burgess, S.~P. Patil, and M.~Trott, {\it {On the Predictiveness of
  Single-Field Inflationary Models}},  {\em JHEP} {\bf 06} (2014) 010,
  [\href{http://arxiv.org/abs/1402.1476}{{\tt arXiv:1402.1476}}].

\bibitem{Bezrukov:2014ipa}
F.~Bezrukov, J.~Rubio, and M.~Shaposhnikov, {\it {Living beyond the edge: Higgs
  inflation and vacuum metastability}},  {\em Phys. Rev.} {\bf D92} (2015),
  no.~8 083512, [\href{http://arxiv.org/abs/1412.3811}{{\tt arXiv:1412.3811}}].

\bibitem{Fumagalli:2016lls}
J.~Fumagalli and M.~Postma, {\it {UV (in)sensitivity of Higgs inflation}},
  {\em JHEP} {\bf 05} (2016) 049, [\href{http://arxiv.org/abs/1602.07234}{{\tt
  arXiv:1602.07234}}].

\bibitem{Enckell:2016xse}
V.-M. Enckell, K.~Enqvist, and S.~Nurmi, {\it {Observational signatures of
  Higgs inflation}},  \href{http://arxiv.org/abs/1603.07572}{{\tt
  arXiv:1603.07572}}.

\bibitem{Ade:2015lrj}
{\bf Planck} Collaboration, P.~A.~R. Ade et~al., {\it {Planck 2015 results. XX.
  Constraints on inflation}},  \href{http://arxiv.org/abs/1502.02114}{{\tt
  arXiv:1502.02114}}.

\bibitem{Lerner:2011ge}
R.~N. Lerner and J.~McDonald, {\it {Distinguishing Higgs inflation and its
  variants}},  {\em Phys. Rev.} {\bf D83} (2011) 123522,
  [\href{http://arxiv.org/abs/1104.2468}{{\tt arXiv:1104.2468}}].

\bibitem{Bezrukov:2008ut}
F.~Bezrukov, D.~Gorbunov, and M.~Shaposhnikov, {\it {On initial conditions for
  the Hot Big Bang}},  {\em JCAP} {\bf 0906} (2009) 029,
  [\href{http://arxiv.org/abs/0812.3622}{{\tt arXiv:0812.3622}}].

\bibitem{Nurmi:2015ema}
S.~Nurmi, T.~Tenkanen, and K.~Tuominen, {\it {Inflationary Imprints on Dark
  Matter}},  {\em JCAP} {\bf 1511} (2015), no.~11 001,
  [\href{http://arxiv.org/abs/1506.04048}{{\tt arXiv:1506.04048}}].

\bibitem{Kainulainen:2016vzv}
K.~Kainulainen, S.~Nurmi, T.~Tenkanen, K.~Tuominen, and V.~Vaskonen, {\it
  {Isocurvature Constraints on Portal Couplings}},  {\em JCAP} {\bf 1606}
  (2016), no.~06 022, [\href{http://arxiv.org/abs/1601.07733}{{\tt
  arXiv:1601.07733}}].

\bibitem{Kainulainen:2015sva}
K.~Kainulainen, K.~Tuominen, and V.~Vaskonen, {\it {Self-interacting dark
  matter and cosmology of a light scalar mediator}},  {\em Phys. Rev.} {\bf
  D93} (2016), no.~1 015016, [\href{http://arxiv.org/abs/1507.04931}{{\tt
  arXiv:1507.04931}}].

\bibitem{Cline:2013gha}
J.~M. Cline, K.~Kainulainen, P.~Scott, and C.~Weniger, {\it {Update on scalar
  singlet dark matter}},  {\em Phys. Rev.} {\bf D88} (2013) 055025,
  [\href{http://arxiv.org/abs/1306.4710}{{\tt arXiv:1306.4710}}]. [Erratum:
  Phys. Rev.D92,no.3,039906(2015)].

\bibitem{Aaltonen:2013ioz}
{\bf CDF, D0} Collaboration, T.~Aaltonen et~al., {\it {Higgs Boson Studies at
  the Tevatron}},  {\em Phys. Rev.} {\bf D88} (2013), no.~5 052014,
  [\href{http://arxiv.org/abs/1303.6346}{{\tt arXiv:1303.6346}}].

\bibitem{Khachatryan:2014jba}
{\bf CMS} Collaboration, V.~Khachatryan et~al., {\it {Precise determination of
  the mass of the Higgs boson and tests of compatibility of its couplings with
  the standard model predictions using proton collisions at 7 and 8 $\,\text
  {TeV}$}},  {\em Eur. Phys. J.} {\bf C75} (2015), no.~5 212,
  [\href{http://arxiv.org/abs/1412.8662}{{\tt arXiv:1412.8662}}].

\bibitem{Aad:2015gba}
{\bf ATLAS} Collaboration, G.~Aad et~al., {\it {Measurements of the Higgs boson
  production and decay rates and coupling strengths using pp collision data at
  $\sqrt{s}=7$ and 8 TeV in the ATLAS experiment}},  {\em Eur. Phys. J.} {\bf
  C76} (2016), no.~1 6, [\href{http://arxiv.org/abs/1507.04548}{{\tt
  arXiv:1507.04548}}].

\bibitem{Akerib:2015rjg}
{\bf LUX} Collaboration, D.~S. Akerib et~al., {\it {Improved Limits on
  Scattering of Weakly Interacting Massive Particles from Reanalysis of 2013
  LUX Data}},  {\em Phys. Rev. Lett.} {\bf 116} (2016), no.~16 161301,
  [\href{http://arxiv.org/abs/1512.03506}{{\tt arXiv:1512.03506}}].

\bibitem{Agashe:2014kda}
{\bf Particle Data Group} Collaboration, K.~A. Olive et~al., {\it {Review of
  Particle Physics}},  {\em Chin. Phys.} {\bf C38} (2014) 090001.

\bibitem{DeSimone:2008ei}
A.~De~Simone, M.~P. Hertzberg, and F.~Wilczek, {\it {Running Inflation in the
  Standard Model}},  {\em Phys. Lett.} {\bf B678} (2009) 1--8,
  [\href{http://arxiv.org/abs/0812.4946}{{\tt arXiv:0812.4946}}].

\bibitem{Moss:2015gua}
I.~G. Moss, {\it {Vacuum stability and the scaling behaviour of the
  Higgs-curvature coupling}},  \href{http://arxiv.org/abs/1509.03554}{{\tt
  arXiv:1509.03554}}.

\bibitem{Buttazzo:2013uya}
D.~Buttazzo, G.~Degrassi, P.~P. Giardino, G.~F. Giudice, F.~Sala, A.~Salvio,
  and A.~Strumia, {\it {Investigating the near-criticality of the Higgs
  boson}},  {\em JHEP} {\bf 12} (2013) 089,
  [\href{http://arxiv.org/abs/1307.3536}{{\tt arXiv:1307.3536}}].

\end{thebibliography}\endgroup

\end{document}